\documentclass[a4paper,11pt]{article}
\usepackage{pos}

\title{The SPD project at NICA}

\author[a]{A. Guskov on behalf of the SPD collaboration}

\affiliation[a]{Joint Institute for Nuclear Research,\\
  Joliot Curie, 6, Dubna, Russia}

\emailAdd{avg@jinr.int}

\abstract{The Spin Physics Detector (SPD) is a universal detector in the one of two interaction points of the NICA collider under construction at JINR, Dubna. SPD plans to study the spin structure of the proton and deuteron and other spin-related phenomena using a unique possibility to operate with polarized proton and deuteron beams at a collision energy up to 27 GeV and a luminosity up to $10^{32}$ cm$^{-2}$ s$^{-1}$. As the main goal, the experiment aims to provide access to the gluon TMD PDFs in the proton and deuteron, as well as the gluon transversity distribution and tensor PDFs in the deuteron, via the measurements of specific single and double spin asymmetries using different complementary probes such as charmonia, open charm, and prompt photon production processes. Other polarized and unpolarized physics is possible, especially at the first stage of NICA operation with reduced luminosity and collision energy of the proton and ion beams. Construction of the first stage of the SPD facility is included in the JINR seven-year development plan for 2024-2030. The physics program of the SPD project and the design of the SPD setup are presented.}

\begin{document}
\maketitle

\section{Introduction}
The spin of the nucleon is defined by its constituents, quarks and gluons, and is built up from their intrinsic spin and orbital angular momenta. The EMC experiment at CERN using a muon beam revealed that quark spins carry only a small fraction  of the proton's spin, triggering decades of research into the origin of proton spin. This research line was continued in numerous experiments at SLAC in the 1990s.
 Initial constraints on the gluon spin contribution (the gluon helicity) came from the fixed-target experiments SMC (CERN), COMPASS (CERN) and HERMES (DESY). They used polarized lepton beams on polarized targets, providing the first significant (though with large uncertainties) evidence for a positive gluon contribution. The HERMES experiment also provided the first detailed data on the transverse momentum distribution of quarks (TMD PDFs) and made a significant contribution to understanding the 3D structure of the proton. These investigations were continued on a larger scale at the COMPASS facility. Important information about the 3D picture of the proton within the complementary approach of generalized parton distributions (GPDs) was also obtained in experiments at JLab, CERN, and DESY. The Relativistic Heavy Ion Collider (RHIC) at BNL is the world's first and only polarized proton-proton collider. Its experiments (STAR and PHENIX) directly measured the gluon helicity contribution to the proton spin and made pioneering measurements of transverse spin asymmetries, providing access to TMD PDFs like the Sivers and Collins functions.
	
	The next-generation facilities are expected to advance toward 3D tomography of the proton's structure in the next decade. The Electron-Ion Collider, EIC (USA) is a future precision facility designed to definitively map the contributions of quark and gluon spins and orbital angular momenta, and perform detailed 3D imaging of the proton and nuclei especially in the range of small Bjorken-x. A similar machine, EIcC at a lower energy range that would be ideal for studying the transition region of QCD, is proposed in China. The Nuclotron-based Ion Collider fAcility (NICA) complex at JINR (Dubna) offers a unique program for spin physics through its dedicated Spin Physics Detector (SPD). Designed to collide polarized protons and deuterons, NICA should provide possibility to study the spin-dependent effects in nuclear collisions in the wide energy range up to $\sqrt{s}=27$ GeV. The main focus of the SPD physics program will be on the gluon Imaging: investigation of gluon TMDs via measurements of specific final states. NICA is expected to remain the world's only polarized hadron collider for the foreseeable future.


 The history of physics with polarized beams at JINR is a long one. In 1976, polarized deuteron beams with energies up to 4.5 GeV/nucleon were obtained for the first time at the Synchrophasotron accelerator. Over 50 years, a series of experiments with polarized proton, deuteron, and neutron beams was carried out at the Synchrophasotron and later at its successor, the Nuclotron. A study of spin effects in N-N scattering and the spin structure of light nuclei in the intermediate energy region was performed. The NICA accelerator complex, now coming online, combined with the significant scientific groundwork and established traditions, will allow spin physics at JINR to reach a qualitatively new level. As the capabilities of the new accelerator grow, the focus will shift from the physics of the intermediate energy region to higher energies for studying the spin-dependent partonic structure of the proton and deuteron.

\section{NICA facility}
\begin{figure}[!h]
  \begin{center}
    \includegraphics[width=1.\textwidth]{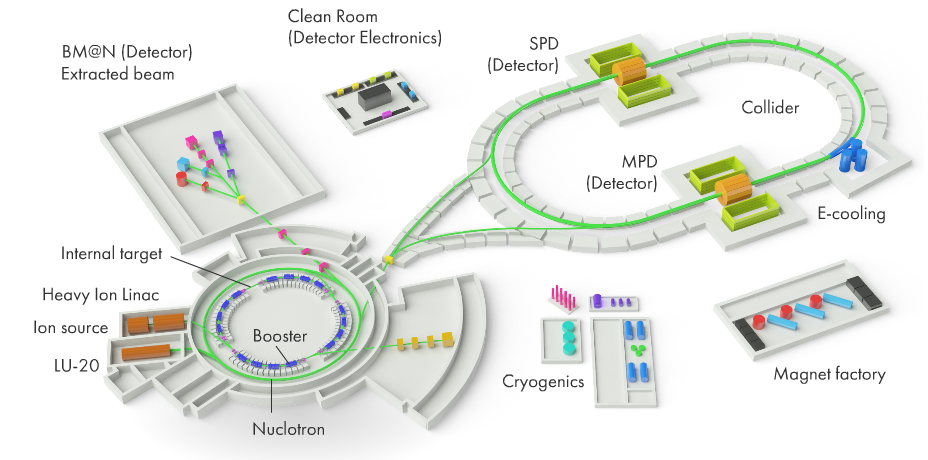}
  \end{center}
  \caption{NICA complex at JINR.}
  \label{fig_NICA}
\end{figure}

The Nuclotron-based Ion Collider fAcility (NICA) is a modern research complex which is currently being put into operation at the Joint Institute for Nuclear Research \cite{Syresin:2021kiy, Lebedev:2023rr}. It is a flagship project dedicated to study the fundamental properties of the strong interaction. The core of NICA complex is the Nuclotron, a superconducting ion synchrotron which has been in operation since 1993. This facility will feature two injection chains: one for heavy ions, including a booster — a compact superconducting synchrotron — and another for light polarized nuclei - protons and deuterons. Additionally, a 503-meters long storage ring is  constructed, where particle collisions are  planned in two interaction points. The main elements of the NICA facility are presented in Fig. \ref{fig_NICA}.


NICA will provide a diverse range of heavy-ion beams up to $Au^{79+}$ ions with a kinetic energy of up to 4.5 GeV/u for heavy-ion physics to be studied at the Multi-Purpose Detector (MPD) at the first interaction point. High-intensity polarized proton beams with a high degree of longitudinal or transverse polarization, a total energy of up to 13.5 GeV and luminosity up to $10^{32}$ cm$^{-2}$ s$^{-1}$ will be used for spin physics studies at the Spin Physics Detector (SPD) which is under construction in the second interaction point. Tensor polarization will also be accessible for deuteron beams operating at half the energy per nucleon and with deuteron–deuteron collision luminosity reduced by an order of magnitude.



\section{Physics program}
\begin{figure}[!h]
  \begin{center}
    \includegraphics[width=1.\textwidth]{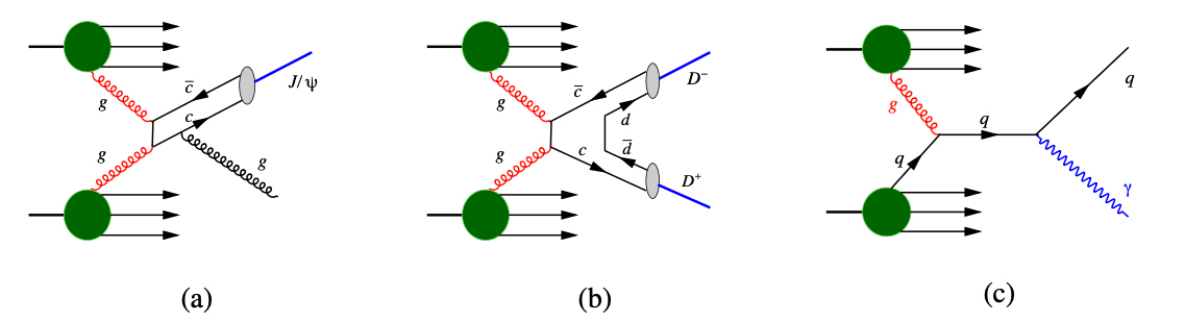}
  \end{center}
  \caption{Partonic-level diagrams illustrating  production of (a) charmonium, (b) open charm, and (c) prompt photons..}
  \label{SPD-probes}
\end{figure}

\begin{figure}[!h]
  \begin{center}
    \includegraphics[width=1.\textwidth]{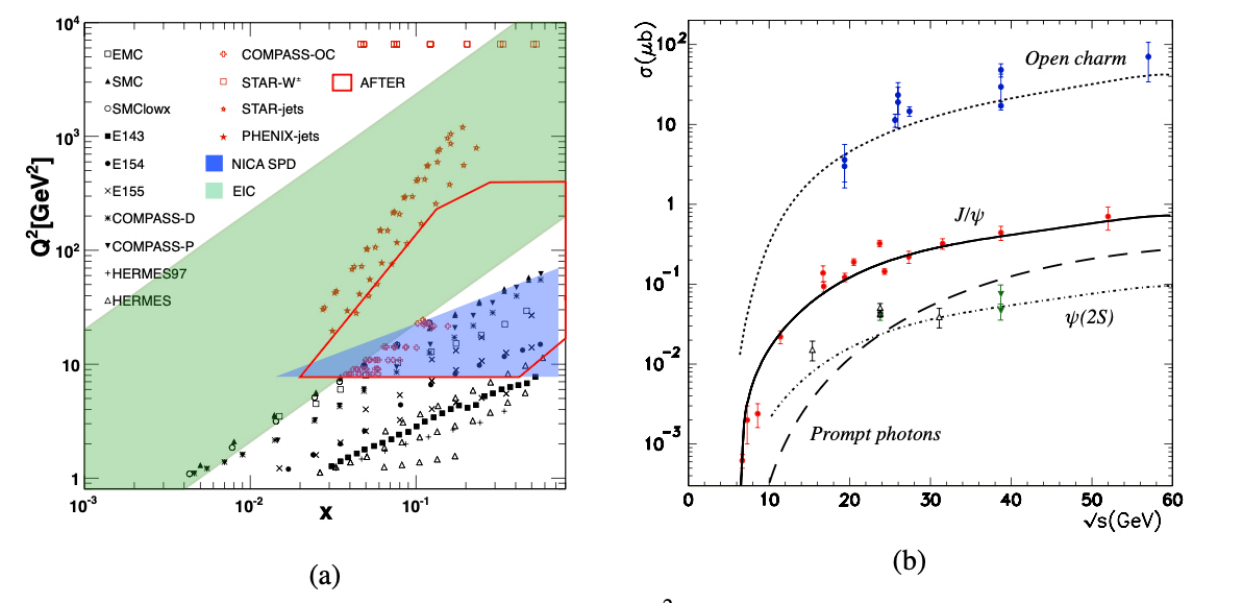}
  \end{center}
  \caption{(a)  The kinematic coverage in the ($x$-$Q^2$) plane for present and future polarized experiments. The domain
     expected to be covered by NICA SPD by the main gluon probes is shown in blue. (b) Cross-sections for the processes of open charm, $J/\psi$, $\psi(2S)$ and prompt photons ($p_T > 3$ GeV/$c$) production as a function of $\sqrt{s}$.}
  \label{SPD-sec}
\end{figure}

The polarized gluon content of proton and deuteron at intermediate and high values of the Bjorken $x$ is the main goal of the Spin Physics Detector research program. Three complementary probes will be used to access it: inclusive production of charmonia, open charm, and high-$p_T$ prompt photons. Corresponding diagrams are shown at Fig. \ref{SPD-probes}. The study of these processes is complementary to such proven approaches to access the partonic structure of the nucleon in hadronic collisions as the inclusive production of hadrons with high transverse momentum and the Drell-Yan process as well as the deep inelastic scattering. The $x$-$Q^2$ kinematic domain to be covered by SPD and energy dependence of the cross-sections for the above-mentioned processes are shown in Fig. \ref{SPD-sec} (a) and (b). Such gluon probes as inclusive productions of neutral and charged pions and
other light mesons, for which the $qg\to qg$ hard process dominates in a certain kinematic region, can also be used to access the polarized gluons at SPD.  Registration of these processes does not impose
additional specific requirements on the experimental setup and can be performed in parallel with the
aforementioned main probes. The same is largely true for the processes that probe quark distributions.

Even in the case of unpolarized collisions, SPD data will be in high demand. The available data constrain weakly the unpolarized gluon density in the proton for $x$ greater than 0.5.  As for the deuteron, there is a prediction that due to nuclear effects in the high‑$x$ region it contains considerably more gluons than the proton and neutron taken separately \cite{Brodsky:2018zdh}.

Collision of the longitudinally polarized beams leads one to the measurement of the $A_{LL}$ asymmetry which is connected to the gluon helicity function $\Delta g(x)$. This asymmetry for the gluon fusion hard processes is sensitive quadratically to $\Delta g(x)$ while the prompt-photon production can be used for determination of the sign of the gluon helicity. The detailed discussion of the expected impact of the SPD measurements to the global knowledge of $\Delta g(x)$ can be found at \cite{Guskov:2023dlj} .

A remarkable difference of the parton transverse distributions from the longitudinal ones is that the gluon transversity distribution does not exist in the spin-1/2 nucleons. Since proton and neutron in the weakly bound deuteron  themselves do not contain the gluon transversity, one could expect vanishing gluon transversity in the deuteron. However, if sizable values for the gluon transversity are measured, this might indicate the presence of new degrees of freedom in the deuteron.

One of the promising approaches to investigate the 3D spin structure of the nucleon is the study of transverse single-spin asymmetries (SSAs) in the interaction of a transversely polarized beam with an unpolarized one. These asymmetries can be used for extraction of the set of the TMD PDFs. Some of these functions are rather well known for quarks, but there is insufficient experimental information to extract the corresponding functions for gluons. The predictions for the Sivers asymmetries, related to the gluon Sivers function, at the SPD energies can be found at \cite{Arbuzov:2020cqg,Saleev:2024unj,Karpishkov:2022ktk,Saleev:2022jda}.

The tensor polarization of the deuteron gives rise to a new set of tensor distributions. To date, only their linear combination for quarks -- the structure function $b_1$ -- has been measured experimentally \cite{HERMES:2005pon}. The possibility of colliding tensor‑polarized deuterons at high energies potentially makes SPD a unique laboratory for a comprehensive study of the deuteron’s tensor structure.

\begin{figure}[!h]
  \begin{center}
    \includegraphics[width=1.\textwidth]{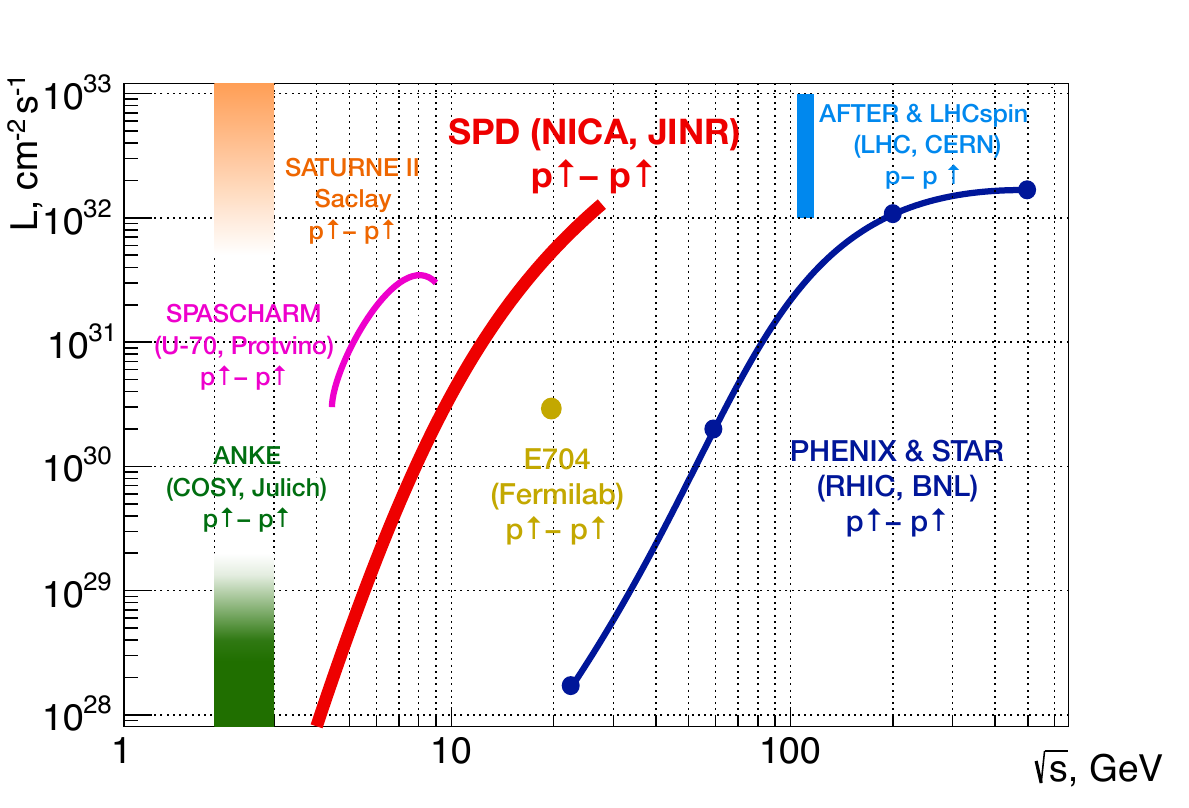}
  \end{center}
  \caption{Luminosity of polarized hadronic collisions as a function of the center-of-mass energy for the  past, present and future general-purpose experiments.}
  \label{SPD-Lumi}
\end{figure}

The SPD project at NICA does not merely fill the energy gap between low-energy experiments with polarized proton beams and RHIC like it is shown in Fig. \ref{SPD-Lumi}. 
Unique to the SPD experiment is the ability to study the same phenomena over a center-of-mass energy range spanning from a few GeV, that corresponds to the non‑perturbative QCD regime dominated by phenomenological models,  up to 27 GeV, where the factorization theorem becomes applicable and hard processes can be calculated within perturbative QCD. Thus, SPD can directly determine the limits of applicability of various factorization approaches via an energy scan.

The first phase of the SPD experiment will be devoted to the study of polarized and non-polarized phenomena at low energies and reduced luminosity using  polarized proton and deuteron beams and heavy ion (up to Ca) beams.
The dedicated physics program of the first stage includes such topics as polarized phenomena in elastic $p$-$p$ and $d$-$d$ scattering and other exclusive reactions, spin effects and correlations in hyperon production, production of multiquark resonances  and hypernuclei, multipartonic correlations, near-threshold charmonia production, etc. \cite{Abramov:2021vtu}. 

\section{SPD experimental setup}
Three main probes have shaped the overall concept of the experimental apparatus.
   The SPD setup is designed as a universal $4\pi$ detector with advanced tracking and particle identification capabilities based on modern technologies \cite{SPD:2024gkq}. The Silicon Vertex Detector will provide  resolution for the vertex position on the level of below 100 $\mu$m needed for the reconstruction of secondary vertices of $D$-meson decays. The Straw tube-based Tracking system  placed within a solenoidal magnetic field of a superconductive magnet of up to 1 T at the detector axis should provide the transverse momentum resolution $\sigma_{p_T}/p_T\approx 2\%$ for a particle momentum of 1 GeV/$c$. The Time-of-Flight system with a time resolution of about 60 ps will provide $3\sigma$ $\pi/K$ and $K/p$ separation of up to about 1.2 GeV/$c$ and 2.2 GeV/$c$, respectively. Possible use of the Focusing Aerogel Ring-Imaging Cherenkov detector (FARICH) in the end-caps will extend this range significantly. Detection of photons will be provided by the sampling Electromagnetic Calorimeter with the energy resolution $\sim 5\%/\sqrt{E}\oplus 1\% $. To minimize multiple scattering and photon conversion effects for photons, the detector material will be kept to a minimum throughout the internal part of the detector. The Range (muon) System  is planned for muon identification. It can also act as a rough hadron calorimeter. The pair of Beam-Beam Counters and Zero Degree Calorimeters will be responsible for the local polarimetry and luminosity control.  To minimize possible systematic effects, the SPD will be equipped with a free-running (triggerless) DAQ system. A high collision rate (up to 4 MHz) and a few hundred thousand detector channels pose a significant challenge to the DAQ,  online monitoring, distributed offline computing system, and data processing software.

 \begin{figure}[!h]
  \begin{center}
    \includegraphics[width=1.\textwidth]{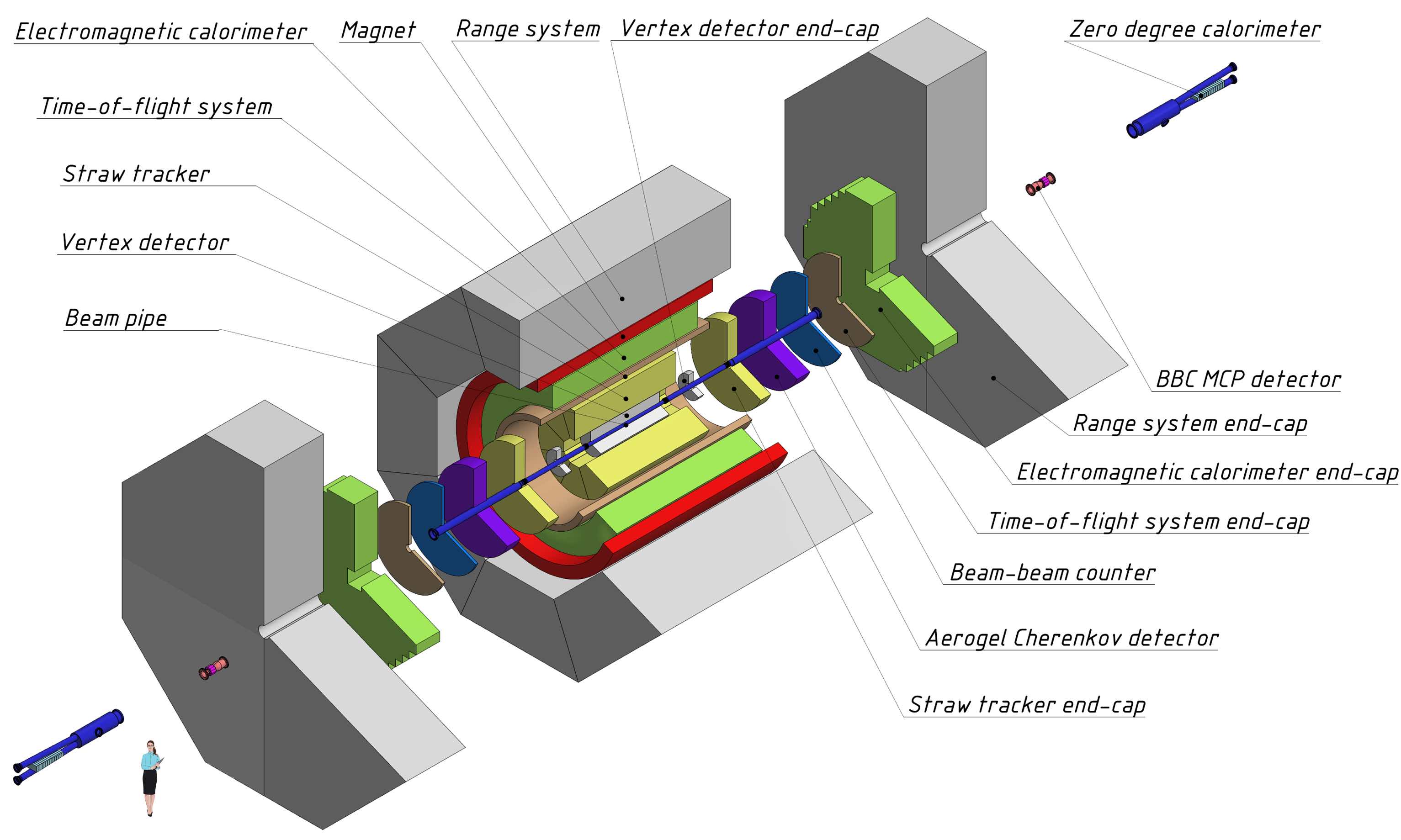}
  \end{center}
  \caption{General layout of the SPD setup.}
  \label{SPD-LA_EX}
\end{figure}

The SPD operation should start a few years after the collider starts operating using the possibilities of polarized $p$-$p$ and $d$-$d$ collisions at $\sqrt{s}<9.4$ GeV and $\sqrt{s}<4.5$ GeV/nucleon, respectively, as well as $A$-$A$ collisions. The starting configuration should consist of the Range System, solenoidal superconducting magnet, Straw tube-based Tracker,  a pair of Zero Degree Calorimeters, and a pair of Beam-Beam Counters. A simple Micromegas-based Central Tracker will be installed in the central region instead of the sophisticated silicone vertex detector to keep a reasonable momentum resolution. Partial installation of the Electromagnetic calorimeter is also planned.

The construction of the initial configuration for the SPD experimental setup is now underway.

 \begin{figure}[!h]
  \begin{center}
    \includegraphics[width=1.\textwidth]{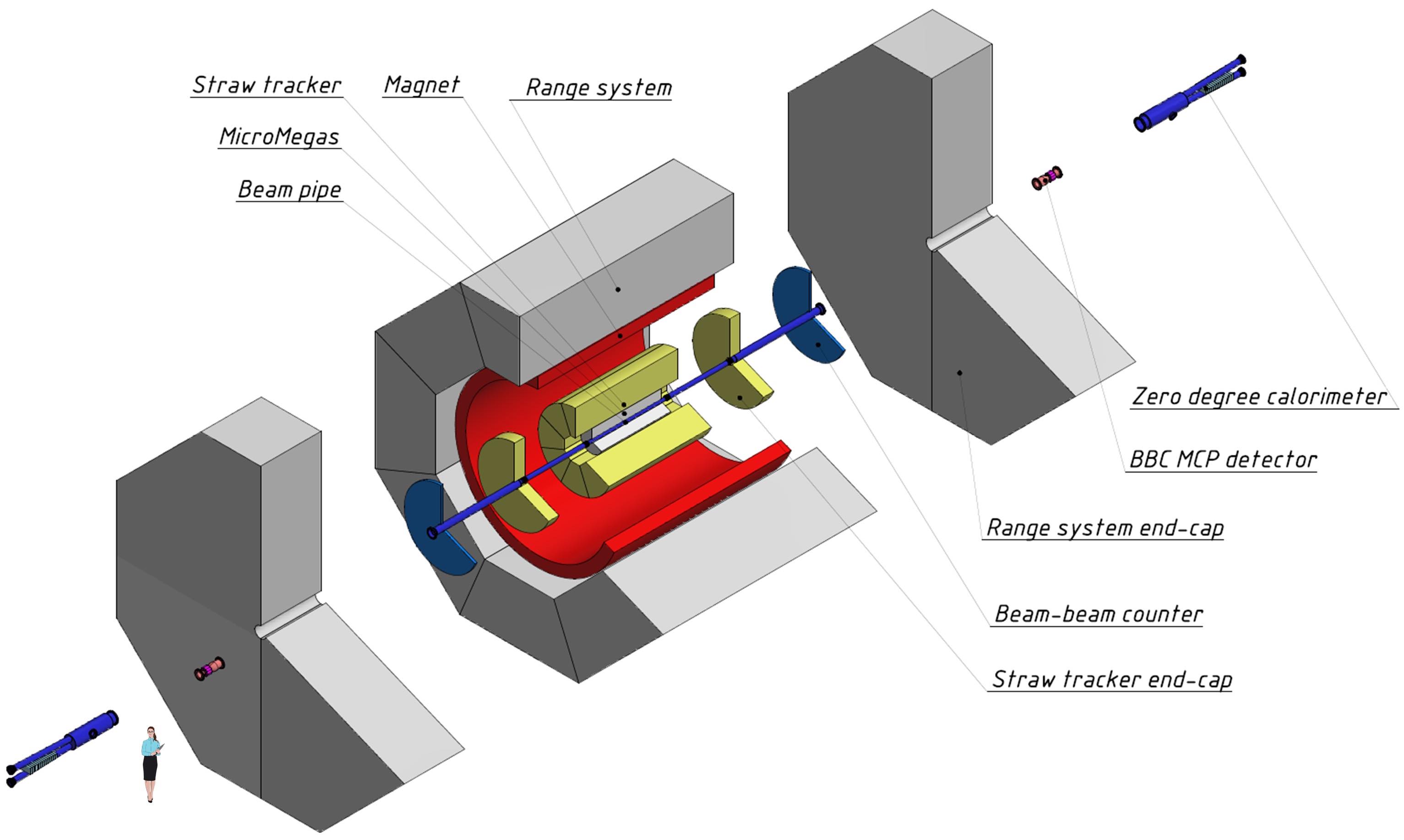}
  \end{center}
  \caption{First-stage layout of the SPD setup.}
  \label{SPD-LA_EX}
\end{figure}

\section{Summary}
The Spin Physics Detector of the NICA collider at JINR is planning as a universal facility for comprehensive study of polarized and unpolarized gluon content of proton and deuteron in polarized high-luminosity $p$-$p$ and $d$-$d$ collisions at $\sqrt{s}\leq 27$ GeV and luminosity up to $10^{32}$ cm$^{-2}$ s$^{-1}$. Complementary main probes such as charmonia, open charm, and prompt photons will be used for that. SPD will contribute significantly to the investigation of the gluon helicity, gluon-induced TMD effects, unpolarized gluon PDFs at high-$x$ in proton and deuteron, gluon transversity in deuteron, and tensor structure of deuteron. Comprehensive physics program for the first period of data taking with reduced energy and luminosity is also under preparation.  The SPD gluon physics program, the implementation of which is planned for the 2030s, is complementary to the other intentions to study the gluon content of hadrons and nuclei at present and future facilities at CERN, JLab, BNL and other leading research centers worldwide. The SPD International Collaboration  \cite{SPD:www} is open to cooperation proposals.

\end{document}